\documentclass[aps,pra,twocolumn,showkeys,showpacs]{revtex4}

\usepackage{amsmath}
\usepackage{amssymb}
\usepackage{dcolumn}

\begin{document}
 
\title{Superdense coding using multipartite states}

\author{Gustavo Rigolin}
\email{rigolin@ifi.unicamp.br}
\affiliation{Departamento de Raios C\'osmicos e Cronologia, Instituto de F\'{\i}sica Gleb Wataghin, Universidade Estadual de Campinas, C.P. 6165, cep 13084-971, Campinas, S\~ao Paulo, Brazil}

\begin{abstract}
We show that with the fourpartite quantum channel used to teleport an arbitrary two qubit state, we can construct a superdense coding protocol where it is possible to transmit $4$ bits of classical information sending only $2$ qubits. Alice and Bob initially share a four qubit maximally entangled state and by locally manipulating her two qubits Alice can generate $16$ orthogonal maximally entangled states, which are used to encode the message transmitted to Bob. He reads the $4$ bit message by a generalized Bell state measurement. A generalized protocol in which $2N$ bits of classical information is transmitted via $N$ qubits is also presented. We also show that this four($2N$-)partite channel is equivalent to two($N$) Bell states, which proves that we need two($N$) Bell states to teleport a two($N$) qubit system.   
\end{abstract}



\maketitle
  
In a seminal paper, Bennett and Wiesner\cite{bennett} proposed what later became known as the superdense coding protocol. This protocol shows how entangled states can increase the communication capacity of two parties. In an ideal classical channel, the transmission of $2$ bits of information requires the manipulation and transmission of at least two particles or physical entities, which are used to encode the information. This means that if Alice wants to transmit $2$ bits to Bob, she must send two particles to him. But if the two parties share a maximally bipartite entangled state ($1/\sqrt{2}(|00\rangle + |11\rangle)$), Alice can transmit $2$ bits of information manipulating and sending only one qubit\cite{bennett}.

Later it was shown that using a tripartite entangled GHZ state ($1/\sqrt{2}(|000\rangle + |111\rangle))$  it is possible to transmit $3$ bits sending only two qubits\cite{gorbachev,cereceda}. Contrasted with the original proposal, where we have $2$ bits of information per transmitted particle, we now have only $1.5$ bit of information per particle.

Here we explicitly show a protocol which enables Alice to transmit $4$ bits of information to Bob sending only two qubits. For this purpose, they must share a four qubit maximally entangled state, in which two qubits are with Alice and the other two are with Bob. By locally operating on her two qubits, Alice can prepare $16$ orthogonal maximally entangled states. By receiving these two qubits Bob can make a generalized Bell state measurement in the four qubits, which enables him to read the $4$ bit message. 

We also prove that this four qubit state is equivalent to a pair of Bell states shared between Alice and Bob. Moreover, since this four qubit state can be used to implement the teleportation of an arbitrary two qubit state $|\Psi\rangle$ $=$ $a|00\rangle$ $+$ $b|01\rangle$ $+$ $c|10\rangle$ $+$ $d|11\rangle$ \cite{rigolin}, this equivalence demonstrates that two Bell states are needed to realize the teleportation of $|\Psi\rangle$. 

Before we present in detail our protocol we briefly review, for clarity purposes, the original superdense coding protocol\cite{bennett}.

Let us assume that Alice and Bob initially share the following Bell state:
\begin{equation}
| \Phi^{+} \rangle = \frac{1}{\sqrt{2}}(| 00 \rangle + | 11 \rangle).
\end{equation}          
Locally operating on her particle, via the following unitary transformations, Alice obtains the four orthogonal Bell states $I | \Phi^{+} \rangle  = | \Phi^{+} \rangle$, $\sigma^{z}_{1} | \Phi^{+} \rangle  =  | \Phi^{-} \rangle$, $\sigma^{x}_{1} | \Phi^{+} \rangle  =  | \Psi^{+} \rangle$, and $i \sigma^{y}_{1} | \Phi^{+} \rangle  =  | \Psi^{-} \rangle$. Here $\sigma$ are the usual Pauli matrices, $I$ is the identity matrix, and the subindex indicates that qubit $1$ is with Alice. The Bell states are defined as
\begin{eqnarray}
| \Phi^{-} \rangle & = & \frac{1}{\sqrt{2}}(| 00 \rangle - | 11 \rangle), \\
| \Psi^{+} \rangle & = & \frac{1}{\sqrt{2}}(| 01 \rangle + | 10 \rangle), \\
| \Psi^{-} \rangle & = & \frac{1}{\sqrt{2}}(| 01 \rangle - | 10 \rangle).
\end{eqnarray}
After implementing one of the four local unitary transformations Alice sends her qubit to Bob. By making a Bell measurement he is able to read the $2$ bit message.       

We now pass to a detailed description of our protocol. First, Alice and Bob must share the following maximally entangled four qubit state:
\begin{equation}
|g_{1}\rangle = \frac{1}{2}\left( |0000\rangle +|0101\rangle + |1010\rangle +|1111\rangle \right).
\end{equation}
The first two qubits belong to Alice and the last two belong to Bob ($|AABB\rangle$, $A \rightarrow$ Alice and $B \rightarrow$ Bob). In order to faithfully transmit a $4$ bit message to Bob, Alice should be able to locally transform state $|g_{1}\rangle$ into the following $16$ maximally entangled states, which we separate in four groups. 

\noindent Group $1$:
\begin{eqnarray}
|g_{1}\rangle & = & \frac{1}{2}\left( |0000\rangle +|0101\rangle + |1010\rangle +|1111\rangle \right), \\
|g_{2}\rangle & = & \frac{1}{2}\left( |0000\rangle +|0101\rangle - |1010\rangle - |1111\rangle \right), \\
|g_{3}\rangle & = & \frac{1}{2}\left( |0000\rangle -|0101\rangle + |1010\rangle - |1111\rangle \right), \\
|g_{4}\rangle & = & \frac{1}{2}\left( |0000\rangle -|0101\rangle - |1010\rangle + |1111\rangle \right).
\end{eqnarray}
Group $2$:
\begin{eqnarray}
|g_{5}\rangle & = & \frac{1}{2}\left( |0001\rangle +|0100\rangle + |1011\rangle + |1110\rangle \right), \\
|g_{6}\rangle & = & \frac{1}{2}\left( |0001\rangle +|0100\rangle - |1011\rangle - |1110\rangle \right), \\
|g_{7}\rangle & = & \frac{1}{2}\left( |0001\rangle -|0100\rangle + |1011\rangle - |1110\rangle \right), \\
|g_{8}\rangle & = & \frac{1}{2}\left( |0001\rangle -|0100\rangle - |1011\rangle + |1110\rangle \right).
\end{eqnarray}
Group $3$:
\begin{eqnarray}
|g_{9}\rangle & = & \frac{1}{2}\left( |0010\rangle +|0111\rangle + |1000\rangle + |1101\rangle \right), \\
|g_{10}\rangle & = & \frac{1}{2}\left( |0010\rangle +|0111\rangle - |1000\rangle - |1101\rangle \right), \\
|g_{11}\rangle & = & \frac{1}{2}\left( |0010\rangle -|0111\rangle + |1000\rangle - |1101\rangle \right), \\
|g_{12}\rangle & = & \frac{1}{2}\left( |0010\rangle -|0111\rangle - |1000\rangle + |1101\rangle \right).
\end{eqnarray}
Group $4$:
\begin{eqnarray}
|g_{13}\rangle & = & \frac{1}{2}\left( |0011\rangle +|0110\rangle + |1001\rangle + |1100\rangle \right), \\
|g_{14}\rangle & = & \frac{1}{2}\left( |0011\rangle +|0110\rangle - |1001\rangle - |1100\rangle \right), \\
|g_{15}\rangle & = & \frac{1}{2}\left( |0011\rangle -|0110\rangle + |1001\rangle - |1100\rangle \right), \\
|g_{16}\rangle & = & \frac{1}{2}\left( |0011\rangle -|0110\rangle - |1001\rangle + |1100\rangle \right).
\end{eqnarray}
These $16$ states form an orthonormal basis: $\sum_{j=1}^{16}|g_{j}\rangle\langle g_{j}|$ $=$ $I$ and $\langle g_{j}|g_{k} \rangle$ $=$ $\delta_{jk}$. From now on, we call it generalized Bell basis. Table \ref{tabela1} shows how Alice can locally operate on her two qubits in order to generate the $16$ generalized Bell states. 
\begin{table}[ht]
\caption{\label{tabela1} This table presents the $16$ local operations Alice should perform on her two qubits in order to obtain the generalized Bell states. These states are used to encode the $4$ bit message Alice transmits to Bob.}
\begin{ruledtabular}
\begin{tabular}{lll}
 Group 1 & Group 2 \\ \hline
$|g_{1} \rangle   =  I |g_{1} \rangle$ & $|g_{5} \rangle   =  \sigma_{2}^{x} |g_{1} \rangle$  \\
$|g_{2} \rangle   =  \sigma_{1}^{z} |g_{1} \rangle$ & $|g_{6} \rangle   =  \sigma_{1}^{z}\sigma_{2}^{x} |g_{1} \rangle$  \\
$|g_{3} \rangle   =  \sigma_{2}^{z} |g_{1} \rangle$ & $|g_{7} \rangle   =  \sigma_{2}^{z} \sigma_{2}^{x}|g_{1} \rangle$  \\
$|g_{4} \rangle   =  \sigma_{2}^{z} \sigma_{1}^{z} |g_{1} \rangle$ & $|g_{8} \rangle   =  \sigma_{2}^{z} \sigma_{1}^{z}\sigma_{2}^{x} |g_{1} \rangle$ \\ \hline \hline
Group 3 & Group 4 \\ \hline 
$|g_{9} \rangle  =   \sigma_{1}^{x} |g_{1} \rangle$ & $|g_{13} \rangle  =   \sigma_{2}^{x}\sigma_{1}^{x} |g_{1} \rangle$ \\
$|g_{10} \rangle  =  \sigma_{1}^{z}\sigma_{1}^{x} |g_{1} \rangle$ & $|g_{14} \rangle  =  \sigma_{1}^{z}\sigma_{2}^{x}\sigma_{1}^{x} |g_{1} \rangle$ \\
$|g_{11} \rangle  =  \sigma_{2}^{z}\sigma_{1}^{x} |g_{1} \rangle$ & $|g_{15} \rangle  =  \sigma_{2}^{z}\sigma_{2}^{x}\sigma_{1}^{x} |g_{1} \rangle$ \\
$|g_{12} \rangle  =  \sigma_{2}^{z}\sigma_{1}^{z}\sigma_{1}^{x} |g_{1} \rangle$ & $|g_{16} \rangle  =  \sigma_{2}^{z}\sigma_{1}^{z}\sigma_{2}^{x}\sigma_{1}^{x} |g_{1} \rangle$
\end{tabular}
\end{ruledtabular}
\end{table}
The $16$ unitary operations which Alice should apply on her two qubits have an interesting feature. They can all be written as $U = U_{1} \otimes U_{2}$, where $1$ and $2$ refer to the first and second qubit respectively. This means that we only need single qubit gates to implement all the $16$ unitary operations. More elaborated two qubit gates, such as the CNOT gate, is not needed. This fact considerably simplifies future implementations of this protocol.  And more, since in the next paragraphs we show that this protocol cannot have its transmission capacity improved, the fact that $U = U_{1} \otimes U_{2}$ is also in agreement with an important result derived in Ref. \cite{bruss}, where it was shown that Alice does not need to perform global unitary operations to obtain the maximal capacity in a dense coding scheme. 

After choosing one of these $16$ local operations, Alice sends her two qubits to Bob. With the four qubits in his possession, Bob performs a generalized Bell measurement in order to discover which generalized Bell state ($|g_{j}\rangle$) Alice has prepared. In this way, Bob can read the $4$ bit message using, for example, the convention (previously discussed with Alice) depicted in Table \ref{tabela2}: 
\begin{table}[ht]
\caption{\label{tabela2} This table shows the $16$ generalized Bell states and the $4$ bit encoding on which Alice and Bob agree.}
\begin{ruledtabular}
\begin{tabular}{lclc}
 State & Convention & State & Convention \\ \hline
$|g_{1} \rangle$ & $0000$ & $|g_{2} \rangle$ & $0001$ \\
$|g_{3} \rangle$ & $0010$ & $|g_{4} \rangle$ & $0100$ \\ 
$|g_{5} \rangle$ & $1000$ & $|g_{6} \rangle$ & $0011$ \\
$|g_{7} \rangle$ & $0110$ & $|g_{8} \rangle$ & $1100$ \\
$|g_{9} \rangle$ & $0101$ & $|g_{10}\rangle$ & $1001$ \\
$|g_{11}\rangle$ & $1010$ & $|g_{12}\rangle$ & $0111$ \\
$|g_{13}\rangle$ & $1011$ & $|g_{14}\rangle$ & $1101$ \\
$|g_{15}\rangle$ & $1110$ & $|g_{16}\rangle$ & $1111$ 
\end{tabular}
\end{ruledtabular}
\end{table}

As when the original proposal was presented, the technological achievements \cite{zeilinger} required to implement this protocol in real world is beyond present day experimental techniques. Three steps must be accomplished in order to realize this protocol. First, the parties should have a source of four maximally entangled states given by $|g_{1}\rangle$. Second, Alice should be able to implement the $16$ local unitary operations, and finally, Bob should realize the generalized Bell measurement. The first two steps is almost already at hand \cite{4photons,5photons}, but we still need an efficient way to discriminate the $16$ generalized Bell states. 

It is interesting to note that the four qubit GHZ state, $|GHZ\rangle = 1/\sqrt{2}(|0000\rangle + |1111\rangle)$, cannot be used to implement the previous protocol. Manipulating her two qubits, Alice can produce only the following eight orthogonal entangled states: 
\begin{eqnarray}
|GHZ^{\pm}\rangle & = & \frac{1}{\sqrt{2}}(|0000\rangle \pm |1111\rangle), \\
|G^{\pm}\rangle & = & \frac{1}{\sqrt{2}}(|0100\rangle \pm |1011\rangle), \\
|H^{\pm}\rangle & = & \frac{1}{\sqrt{2}}(|1000\rangle \pm |0111\rangle), \\
|Z^{\pm}\rangle & = & \frac{1}{\sqrt{2}}(|1100\rangle \pm |0011\rangle).
\end{eqnarray} 
Therefore, using a GHZ state and locally operating on her two qubits, Alice can at most send a $3$ bit message to Bob. This is the justification  for calling the $|g_{j}\rangle$ states \textit{maximally} entangled. At least for this task, the GHZ states are less entangled than the $|g_{j}\rangle$ states. We can also get a physical insight of the communication power of these $|g_{j}\rangle$ states noting that the previous protocol can be adapted to faithfully teleport \cite{lee,rigolin} a two qubit state $|\Psi\rangle$ $=$ $a|00\rangle$ $+$ $b|01\rangle$ $+$ $c|10\rangle$ $+$ $d|11\rangle$. These four qubit states can also be used to define a new multipartite entanglement measure, which has a simple physical interpretation and is easily extended to a $2N$ qubit state. This measure quantifies the usefulness of a quantum state as a quantum teleportation channel \cite{rigolin}. 

This generalized superdense coding protocol is also optimum. By optimum we mean that its capacity to transmit classical bits is equal to the Holevo bound \cite{bruss}. The Holevo bound is the maximal amount of classical information a $d$-dimensional quantum system can transmit: $H =\log_{2}d$ bits. Since the $|g_{j}\rangle$ state has $d = 2^{4}$, $H = 4$ bits, which is exactly the number of transmitted bits of the generalized protocol. In general, the capacity of dense coding $\chi$ for a given shared state $\rho^{AB}$ between two parties $A$ and $B$ is given by \cite{bruss}
\begin{equation}
\chi(\rho^{AB}) = \log_{2} d_{A} + S(\rho^{B}) - S(\rho^{AB}), \label{capacity}
\end{equation}
where $d_{A}$ is the dimension of Alice's system, $\rho^{B} = Tr_{A}(\rho^{AB})$ is the reduced density matrix with respect to subsystem $A$, $\rho^{AB}$ is the system density matrix, and $S(\sigma) = -Tr(\sigma\log_{2}\sigma)$ is the von Neumann entropy. Using Eq.~(\ref{capacity}) we can again see that the generalized protocol achieves the Holevo bound. For $\rho^{AB} = |g_{1}\rangle\langle g_{1}|$ we have $d_{A} = 4$ and a simple calculation shows that $S(\rho^{B}) = 2$ and $S(\rho^{AB}) = 0$. Using these results in Eq.~(\ref{capacity}) we get $\chi(|g_{1}\rangle) = 4$. Repeating the above calculations for the GHZ state we obtain $\chi(|GHZ\rangle) = 3$. Looking at these two capacities of dense coding it is clear that we definitely cannot achieve the Holevo bound using a GHZ state and we should use a $g_{1}$ state to maximize our channel capacity. 
  
The previous protocol can also be generalized to an even number of qubits. If Alice and Bob wish to transmit $2 N$ bits of information, they will need to share a $2 N$ multipartite maximally entangled state and Alice will need to send $N$ particles. At the end of the protocol, they will always obtain a $2$ bit rate of useful information per transmitted particle. The superdense coding protocol can be rigorously generalized as follows: (a) The generalized Bell state $|g_{1}\rangle$ is written as $|s_{0}\rangle$ $=$ $(2^{-N/2})$ $\sum_{j=0}^{M}$ $|x_{j}\rangle_{A}$ $|x_{j}\rangle_{B}$, where $M=2^{N}-1$ and $x_{j}$ is the binary representation of the number $j$. In the $4$ bit superdense coding protocol, $x_{0}=00$, $x_{1}=01$, $x_{2}=10$, and $x_{3}=11$. Zeros should be added to make all $x_{j}$ with the same amount of bits ($N$ bits). (b) From $|s_{0}\rangle$ it is possible to obtain all the $2^{2N}$ generalized Bell states locally operating on its first $N$ qubits, $|s_{j}\rangle = \bigotimes_{k=1}^{N}(\sigma_{k}^{z})^{j_{2k-1}}(\sigma_{k}^{x})^{j_{2k}}|s_{0}\rangle$. Now $j_{k}$ represents the $k$-th bit (from right to left) of the number $0 \leq j \leq 2^{2N} - 1$, which is written in binary notation and again zeros should be added to leave all $j$'s with the same number of bits ($2N$ bits). The subindex $k$ indicate on which qubit the Pauli matrices $\sigma^{x}$ and $\sigma^{z}$ should operate. For the $4$ bit protocol shown above these operations are listed in Table \ref{tabela1} and $|s_{0}\rangle = |g_{1}\rangle$, $|s_{1}\rangle = |g_{2}\rangle$, $|s_{2}\rangle = |g_{9}\rangle$, $|s_{3}\rangle = |g_{10}\rangle$, and so on. (c) After implementing one of the $2^{2N}$ local operations on her qubits, Alice sends to Bob the ${N}$ qubits in her possession. (d) With the $2N$ qubit state, Bob makes a generalized Bell measurement reading the $2 N$ bit message.  

Now we prove that the state $|g_{1}\rangle$ is equivalent to a pair of two Bell states. For clarity purposes we explicitly rewrite $|g_{1}\rangle$ specifying which qubits belong to Alice and which to Bob:
\begin{eqnarray}
|g_{1}\rangle  & = & \frac{1}{2}\left( |0_{A}0_{A}0_{B}0_{B}\rangle +|0_{A}1_{A}0_{B}1_{B}\rangle  \right. \nonumber \\
 & & + |1_{A}0_{A}1_{B}0_{B}\rangle \left.+|1_{A}1_{A}1_{B}1_{B}\rangle \right).
\end{eqnarray}
Changing the order we write the second and third qubits we have
\begin{eqnarray}
|g_{1}\rangle  & = & \frac{1}{2}\left( |0_{A}0_{B}0_{A}0_{B}\rangle +|0_{A}0_{B}1_{A}1_{B}\rangle  \right. \nonumber \\
 & & + |1_{A}1_{B}0_{A}0_{B}\rangle \left.+|1_{A}1_{B}1_{A}1_{B}\rangle \right). \nonumber \\
 & = & \frac{1}{\sqrt{2}}(|0_{A}0_{B}\rangle + |1_{A}1_{B}\rangle)\frac{1}{\sqrt{2}}(|0_{A}0_{B}\rangle + |1_{A}1_{B}\rangle) \nonumber \\
& = & |\Phi^{+}\rangle_{AB} |\Phi^{+}\rangle_{AB}. \label{boa}
\end{eqnarray} 
Actually, repeating the procedure leading to Eq.~(\ref{boa}), all the $|g_{j}\rangle$ states can be written as a direct product of two Bell states. These results are summarized in Table \ref{tabela3}. 
\begin{table}[ht]
\caption{\label{tabela3} This table shows the $|g_{j}\rangle$ states and their corresponding decompositions in two Bell states.}
\begin{ruledtabular}
\begin{tabular}{lll}
 Group 1 & Group 2 \\ \hline
$|g_{1} \rangle = |\Phi^{+}\rangle_{AB} |\Phi^{+}\rangle_{AB}$ & $|g_{5} \rangle = |\Phi^{+}\rangle_{AB} |\Psi^{+}\rangle_{AB}$  \\
$|g_{2} \rangle = |\Phi^{-}\rangle_{AB} |\Phi^{+}\rangle_{AB}$ & $|g_{6} \rangle = |\Phi^{-}\rangle_{AB} |\Psi^{+}\rangle_{AB}$  \\
$|g_{3} \rangle = |\Phi^{+}\rangle_{AB} |\Phi^{-}\rangle_{AB}$ & $|g_{7} \rangle = |\Phi^{+}\rangle_{AB} |\Psi^{-}\rangle_{AB}$  \\
$|g_{4} \rangle = |\Phi^{-}\rangle_{AB} |\Phi^{-}\rangle_{AB}$ & $|g_{8} \rangle = |\Phi^{-}\rangle_{AB} |\Psi^{-}\rangle_{AB}$ \\ \hline \hline
Group 3 & Group 4 \\ \hline 
$|g_{9} \rangle = |\Psi^{+}\rangle_{AB} |\Phi^{+}\rangle_{AB}$ & $|g_{13} \rangle = |\Psi^{+}\rangle_{AB} |\Psi^{+}\rangle_{AB}$ \\
$|g_{10} \rangle = |\Psi^{-}\rangle_{AB} |\Phi^{+}\rangle_{AB}$ & $|g_{14} \rangle = |\Psi^{-}\rangle_{AB} |\Psi^{+}\rangle_{AB}$ \\
$|g_{11} \rangle = |\Psi^{+}\rangle_{AB} |\Phi^{-}\rangle_{AB}$ & $|g_{15} \rangle = |\Psi^{+}\rangle_{AB} |\Psi^{-}\rangle_{AB} $ \\
$|g_{12} \rangle = |\Psi^{-}\rangle_{AB} |\Phi^{-}\rangle_{AB}$ & $|g_{16} \rangle = |\Psi^{-}\rangle_{AB} |\Psi^{-}\rangle_{AB}$
\end{tabular}
\end{ruledtabular}
\end{table}

A straightforward generalization of the previous proof shows that the $2N$ qubit state $|s_{0}\rangle$, which is the quantum channel required to teleport a $N$ qubit state\cite{rigolin}, can be written as $|s_{0}\rangle = |\Phi^{+}\rangle_{AB}^{\otimes N}$. This last result just expresses the necessity of $N$ Bell states to teleport a general pure $N$ qubit system. 

This equivalence is an interesting result for two reasons. First, it may open the possibility, at least for tasks as dense coding and teleportation involving only two parties, to quantify multipartite states relating them to bipartite maximally entangled states (Bell states). Second, it is intriguing to note that in order to teleport \cite{rigolin} an arbitrary two qubit state  $|\Psi\rangle$ $=$ $a|00\rangle$ $+$ $b|01\rangle$ $+$ $c|10\rangle$ $+$ $d|11\rangle$ we need two Bell states ($|g_{1}\rangle$) and not `true' fourpartite entangled states such as $|GHZ\rangle = 1/\sqrt{2}(|0000\rangle + |1111\rangle)$. Indeed, it can be shown that a fourpartite GHZ state can teleport only some special classes of two qubits but not a general one\cite{rigolin}.       

We have shown a superdense coding protocol which permits Alice to faithfully transmit a $4$ bit message to Bob sending only two qubits. For this protocol to work, Alice and Bob should share a generalized four qubit Bell state, Alice should be able to perform local operations on her two qubits, and Bob should be capable of discriminating $16$ orthogonal generalized Bell states. 

We proved that this protocol is optimum because it achieves the Holevo bound. Furthermore, we showed that it is equivalent to two execution of the Bennett and Wiesner original proposal since our fourpartite state was shown to be equivalent to a pair of Bell states. 

The feasibility of an experimental implementation of this protocol and a generalization to multipartite qubits with an even number of constituents were also discussed.   

\begin{acknowledgments}
The author would like to express his gratitude to the funding of Funda\c{c}\~ao de Amparo \`a Pesquisa do Estado de S\~ao Paulo (FAPESP) and  to Dr. Fernando da Rocha Vaz Bandeira de Melo for initiating a discussion which culminated in this work. 
\end{acknowledgments}

\end{document}